# Conductance Switching of Azobenzene-Based Self-Assembled Monolayers on Cobalt Probed by UHV Conductive-AFM.

Louis Thomas,[1,#] Imane Arbouch,[2,#] David Guérin,[1] Xavier Wallart,[1] Colin van Dyck,[2] Thierry Mélin,[1] Jérôme Cornil,[2,*] Dominique Vuillaume[1,*] and Stéphane Lenfant[1,*]

*1) Institute of Electronics Microelectronics and Nanotechnology (IEMN), CNRS, University of Lille, Avenue Poincaré, Villeneuve d'Ascq, France.*

*2) Laboratory for Chemistry of Novel Materials, University of Mons, Place du Parc 20, Mons, Belgium.*

**ABSTRACT**

We report the formation of self-assembled monolayers of a molecular photoswitch (azobenzene-bithiophene derivative, AzBT) on cobalt via a thiol covalent bond. We study the electrical properties of the molecular junctions formed with the tip of a conductive atomic force microscope under ultra-high vacuum. The statistical analysis of the current-voltage curves shows two distinct states of the molecule conductance, suggesting the coexistence of both the *trans* and *cis* azobenzene isomers on the surface. The *cis* isomer population (*trans* isomer) increases (decreases) upon UV light irradiation. The situation is reversed under blue light irradiation. The experiments are confronted to first-principle calculations performed on the molecular junctions with the Non-Equilibrium Green's

Function formalism combined with Density Functional Theory (NEGF/DFT). The theoretical results consider two different molecular orientations for each isomer. Whereas the orientation does not affect the conductance of the *trans* isomer, it significantly modulates the conductance of the *cis* isomer and the resulting conductance ON/OFF ratio of the molecular junction. This helps identifying the molecular orientation at the origin of the observed current differences between the *trans* and *cis* forms. The ON state is associated to the *trans* isomer irrespective of its orientation in the junction, while the OFF state is identified as a *cis* isomer with its azobenzene moiety folded upward with respect to the bithiophene core. The experimental and calculated ON/OFF conductance ratios have a similar order of magnitude. This conductance ratio seems reasonable to make these Co-AzBT molecular junctions a good test-bed to further explore the relationship between the spin-polarized charge transport, the molecule conformation and the molecule-Co spinterface.

**INTRODUCTION**

Molecular spintronics uses molecules and/or ensemble of molecules in devices in which the flow of spin-polarized currents is used to manipulate and process information.[1-4] Prototypical devices include molecular magnetic tunnel junctions where a self-assembled monolayer (SAM) is used as a tunnel layer between two ferromagnetic (FM) electrodes.[5, 6] The study of photo-switchable molecular junctions with FM electrodes is motivated by theoretical works[7-9] predicting that the interplay between molecular conformation and the injection of spin-polarized currents unlocks additional degrees of freedom to control the spin-polarized transport in optically switchable molecular spintronic devices. We have recently studied the optically induced conductance switching at the nanoscale of diarylethene derivative SAMs on $La_{0.7}Sr_{0.3}MnO_3$ electrodes,[10] and observed a weak conductance switching of the diarylethene molecular junctions (ON/OFF conductance ratios <8) coupled to the optically induced resistive switching of the $La_{0.7}Sr_{0.3}MnO_3$ substrate.

Here, we investigate the grafting of thiol-terminated molecular switches on Co surface under controlled atmosphere and the electrical properties of the resulting Co-SAM/metal junctions (- denotes here a covalent binding whereas / denotes a non-covalent binding) formed by conductive atomic force microscopy (CAFM) under ultra-high vacuum (UHV), a mandatory condition to avoid oxidation of the Co electrode. Indeed, despite previous realizations of molecular junctions employing SAMs grafted on cobalt as bottom electrodes,[11-17] electrical nanoscale characterization of SAMs on cobalt remains limited, and photo-switchable molecular junctions on Co are not reported to date.

We study the properties of SAMs composed of a functional molecule, *i.e.* an azobenzene-bithiophene molecular photoswitch (AzBT) (*Scheme 1*)[18] possessing two photo-isomers, *trans*-AzBT and *cis*-AzBT. The conversion from *trans*-AzBT to *cis*-AzBT takes place under UV irradiation whereas the reverse reaction occurs under irradiation with blue light as usual for azobenzene derivatives. Reversible isomerization was also demonstrated by plasmon excitation when AzBT molecules are embedded in a network of gold nanoparticles.[19] We have previously studied the electronic properties of AzBT SAMs on gold by CAFM[20] and in self-assembled networks of AzBT functionalized gold nanoparticles.[19, 21, 22] In all cases, the *trans* and *cis* isomers are respectively associated with the OFF (less conducting) and ON (more conducting) states of the system. The highest ON/OFF ratios (ratio of the currents in the *cis* and *trans* states at a given bias) are ~ 600 in nanoparticles networks[21] and ~ $10^3$ with SAMs.[20]

We first characterize by various techniques (X-ray photoelectron spectroscopy, ellipsometry and topographic AFM) the grafting of the thiol-terminated molecular switches on the surface of cobalt under controlled atmosphere. We investigate the electrical properties (current vs. voltage, I-V, curves) of the resulting Co-SAM/PtIr junctions formed by CAFM under ultra-high vacuum (UHV). We found that, in all situations (pristine SAM, after UV and visible light irradiations), there is a coexistence of the *trans* and *cis* azobenzene isomers in the SAMs. After UV light irradiation, the *cis* isomer population increases while that of the *trans* isomer decreases. The respective populations are reversed upon visible light irradiation. The conductance ratio between the two isomers is about 23 (at 0.5 V).

At the theoretical level, we perform a first-principle investigation of the electronic structure and transport properties of molecular junctions including AzBT in several conformations chemisorbed on the cobalt (111) surface. Since the work function modification upon SAM deposition is related to the charge injection barriers from the metal to the active molecular layer,[23-25] we start our study by characterizing the change in the work function of the cobalt surface upon AzBT deposition and then depict the energy level alignment at the Co-AzBT interface. In a next stage, we simulate the transport properties of Co-AzBT junctions by means of the NEGF/DFT formalism.[26]

The transmission spectra at zero bias are calculated for the various molecular junctions and rationalized by analyzing the nature of the transmitting molecular orbitals through the Local Device Density of States (LDDoS) of the device. The I-V curves are obtained by using the Landauer-Büttiker formalism within the coherent transport regime[27] via integration of the transmission spectrum in the bias window defined by the applied bias. The calculations identify the most probable Co-AzBT interface conformations related to the experimentally observed changes in the molecular junction conductance upon light irradiation. Based on our results, the high conductance ON state corresponds to the *trans* conformation for the Co-AzBT interface irrespective of the orientation of the azobenzene moiety with respect to the bithiophene unit (syn or anti), which gives almost the same conductance. The lower conductance OFF state is associated with one *cis* conformation of the Co-AzBT interface, with the azobenzene moiety folded upward. The lower conductance of the *cis* conformation is attributed to an increase in both the hole injection barrier and tunneling barrier length. We calculate ON/OFF (*trans/cis*) conductance ratios in good agreement with the experimental data.

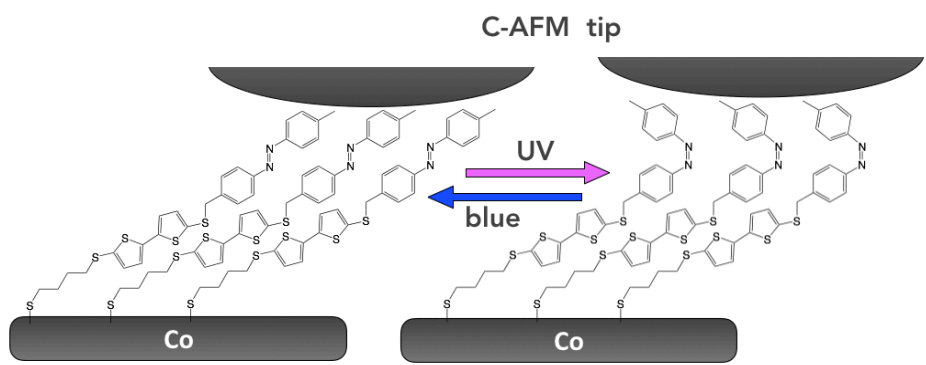

*Scheme 1. Schematic representation of trans azobenzene-bithiophene (trans-AzBT, left) and cis azobenzene-bithiophene (cis-AzBT, right) on Co electrodes connected by the CAFM tip (not to scale).*

## RESULTS

The synthesis of the AzBT molecules was reported elsewhere.[18] The sample fabrication (SAMs on evaporated Co substrates) was done in an inert atmosphere (glove box). The SAMs were formed from a millimolar solution of AzBT in anhydrous toluene in the dark. For the transfer of the samples between the globe box and various instruments (under inert atmosphere or UHV), we used a hermetic container under an inert atmosphere (see Methods).

**Structural characterization of the Co-AzBT SAMs**

***SAM morphology and thickness***. The topographic images of the SAMs acquired simultaneously with the CAFM images (see Methods) reveal a rather homogeneous surface free of pinholes with a rms roughness of ~ 0.3 nm (Fig. 1a), and a spatially uniform distribution of low currents ( ~ few nA at 0.5 V) (Fig. 1b). A first evaluation of the SAM thickness is obtained by the AFM nano-etching technique (see Methods). The topographic images of the indented zone reveal the presence of a square hole with a depth of ~1.1 nm (Figs. 1c and 1e, thickness averaged between the dashed lines in Fig. 1c) where hot spots of current (up to 100 nA - 1 μA at 10 mV) are clearly observed (bright spots in Fig. 1d) as well as a clear increase in the averaged current in the etched zone (Fig. 1f), indicating the removal of the molecular layer. The thickness is confirmed by spectroscopic ellipsometry (see Methods) on a freshly formed Co-AzBT SAM. We measure a thickness of $0.9 \pm 0.2$ nm, in good agreement with the value obtained by nano-etching. The thickness (ellipsometry) increases to $1.5 \pm 0.2$ nm after *trans*-to-*cis* isomerization (UV light irradiation). Considering the calculated (MOPAC, PM3 level)[28] geometry optimized length of the molecules (3.1 nm for the *trans* isomer, 2.6 nm for the *cis* isomer), the average tilt angle of the molecules in the SAM is estimated to $73 \pm 4°$ with respect to the normal of the surface. We conclude that low-density SAMs (about 1.8 nm²/molecule, see theory section in the supporting information) are formed on the Co surface, compared to SAMs of the same molecules on Au (SAM thickness 2.7 - 2.9 nm, with AzBT molecules standing almost upright on the surface).[20] It is likely that despite the strict protocol used (see Methods), a weak oxidation of the Co substrate cannot be avoided (XPS section, *vide infra*), preventing the formation of a densely packed SAM on the surface.

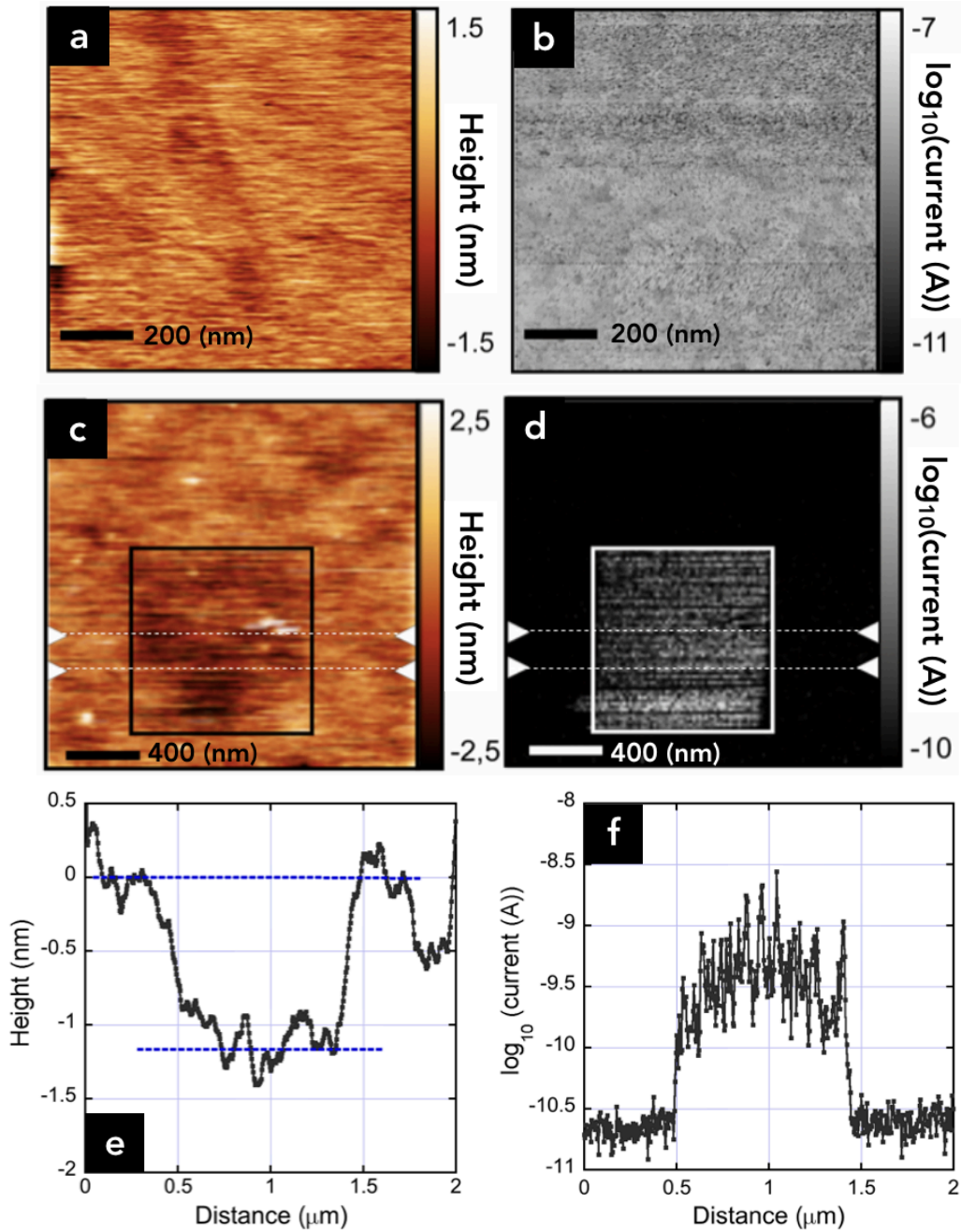

***Figure 1.*** *Topography **(a)** and log(current) **(b)** (1.0×1.0 μm² area, F = 20 nN, V = +0.5 V) on Co-AzBT. Scale bar length is 200 nm. Topography **(c)** and log(current) **(d)** (2.0×2.0 μm², F = 20 nN, V = +0.01 V). The squares indicate the location of a previous scan at F = 140 nN (nano-etching). Scale bar length is 400 nm. **(e, f)** Respective mean sections of (c, d) calculated in the area delimited by triangular cursors and dashed lines.*

***X-ray Photoemission Spectroscopy***. The XPS spectrum of the freshly grafted SAM on cobalt surface (Fig. 2a) shows the presence of: i) the different atoms of the AzBT molecule: carbon (C1s), nitrogen (N1s), sulfur (S2p); ii) cobalt atoms from the substrate (Co2p); and iii) oxygen atoms (O1s). We also record the Co2p, O1s and C1s spectra on an air-exposed cobalt substrate to compare the oxidation state and the carbon contamination on the Co-AzBT sample (see the Supporting Information). The Co2p spectra (Fig. 2b) shows the $2p_{3/2}$ and $2p_{1/2}$ peaks associated with metallic cobalt and cobalt oxides. The first peak at 778.3 eV binding energy is in agreement with the work of Chuang *et al.*[29] where this peak is observed at 778.2 eV and associated to the metallic $Co2p_{3/2}$.[30] The second peak at 780.5 eV corresponds to the binding energy of the cobalt oxides and hydroxides ($CoO$, $Co_3O_4$, $Co_2O_3$ $CoOOH$,...).[29, 30] From the relative amplitudes of the peaks associated to cobalt oxides and hydroxides with respect to the peaks associated to the metallic cobalt, the Co-AzBT sample shows a larger amount of metallic cobalt than the air-exposed sample without SAM (Fig 2b). Albeit not completely suppressing the Co oxidation, the glovebox environment, the protection by the SAM and the hermetic homemade transport container clearly reduce the oxidation of the cobalt surface. The shape of the Co2p spectra is similar to that of Co functionalized under glovebox by highly dense SAMs of hexadecanethiol in the work of Hoertz *et al.*.[12] This presence of a residual oxidized Co in the Co-AzBT is confirmed by the analysis of the O1s spectra (see the Supporting Information).

The C1s spectra (see the Supporting Information) can be deconvoluted in three peaks: the main peak located at 285.1 eV is associated to the C-C and C-S bonds.[31-33] The two other peaks with binding energy 285.6 eV and 288.7 eV correspond to C-O and O-C=O bonds, respectively.[34] These last two peaks are not associated with the AzBT molecule but to a surface contamination. We estimate a ratio of about 80/20 % for the carbon in AzBT and contaminants, respectively (see the Supporting Information).

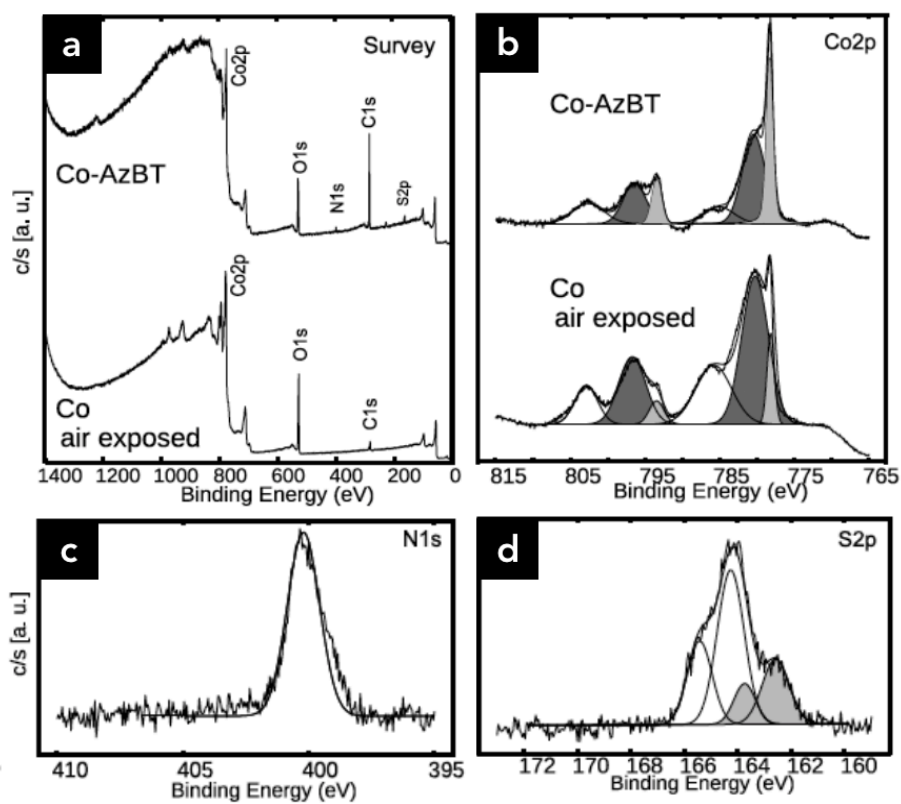

*Figure 2. (a)* XPS survey spectra of Co-AzBT (top) and of an air exposed cobalt substrate (bottom). *(b)* High-resolution spectra of Co2p for Co-AzBT (top) and for air exposed cobalt substrate (bottom). Dark grey peaks are associated to Co oxides and hydroxides and light grey peaks indicate the Co metal. *(c)* High-resolution spectra of N1s for Co-AzBT. *(d)* High-resolution spectra of S2p for Co-AzBT. Light grey peaks are associated to S bound to Co.

The peak position of the nitrogen N1s is observed at 400.2 eV (Fig. 2c), in good agreement with previous results for AzBT grafted on gold substrate where the N1s peak was observed at 399.5 eV.[20] This peak is clearly associated with the presence of the AzBT molecules.

The S2p spectrum (Fig. 2d) shows two doublets corresponding to the spin-orbit splitting of the S2p level into a high intensity $S2p_{3/2}$ peak at lower energy and a low intensity $S2p_{1/2}$ peak at higher energy. This doublet is separated by 1.2 eV with an intensity ratio of 2/1.[18, 33, 35] Two doublets are observed in the S2p spectrum of Co-AzBT SAM. The first doublet, with peaks located at 162.6 eV and 163.7 eV, corresponds to the $S2p_{3/2}$ and $S2p_{1/2}$ respectively. For the second doublet, the $S2p_{3/2}$ peak is located at 165.4 eV and the $S2p_{1/2}$ peak at 164.3 eV. By comparison, XPS of dodecanethiol SAMs formed on cobalt substrate exhibits the $S2p_{3/2}$ component associated to thiolate species centered at 162.2 eV.[14] This value is close to the $S2p_{3/2}$ peak position of the first doublet at 162.6 eV so that doublet may be assigned to the thiol chemisorbed on the cobalt surface ($S2p_{3/2}$-Co). The second doublet corresponds to other sulfur atoms in the body of the molecule ($S2p_{3/2}$-C).

**Electrical characterization by UHV Conductive-AFM.**

The I-V curves are recorded with the UHV CAFM by forming a molecular junction with the PtIr tip at a loading force of ~ 20 nN, resulting in a SAM deformation of ~ 0.2 nm, a contact area of ~17 $nm^2$ , i.e. ~ 10 molecules in the junction (see Methods and more details in the Supporting Information). These curves are acquired just after the fabrication and transfer into the UHV CAFM. Figure 3b shows the 2D I-V histogram of the pristine Co-AzBT/PtIr junctions constructed from 1200 I-V traces (400 I-V acquired on a 20 x 20 grid, pitch = 25 nm, repeated on 3 zones on the SAM, see details in Methods and in the supporting information). These I-V curves have an almost symmetric shape with respect to the bias polarity, as previously observed for the same molecules on Au substrates.[20] The 1D histograms taken at – 0.5 V and + 0.5 V (Figs. 3a and 3c) are decomposed into a high and narrow peak at higher currents (HC peak) and a smaller and broader peak at lower currents (LC peak), which are fitted by two log-normal distributions. Table 1 gives the fitted values of the log-mean current (log-$\mu$) and the log-standard deviation (log-$\sigma$). At both bias polarities, the LC and HC peaks are centered in the nA range and separated by about a half-decade shift (HC peak: log-$\mu$ = -8.38 (4.17 × $10^{-9}$ A), log-$\sigma$ = 0.21 at -0.5 V, log-$\mu$ = -8.42 (3.80 × $10^{-9}$ A), log-$\sigma$ = 0.19 at +0.5 V;

LC peak: log-μ = -8.84 (1.45 × 10$^{-9}$ A), log-σ = 0.40 at -0.5 V, log-μ = -8.89 (1.29 × 10$^{-9}$ A), log-σ = 0.39 at +0.5 V). From the areas of the two peaks, we deduce a relative contribution of 39% for the HC peak and 61% for the LC one.

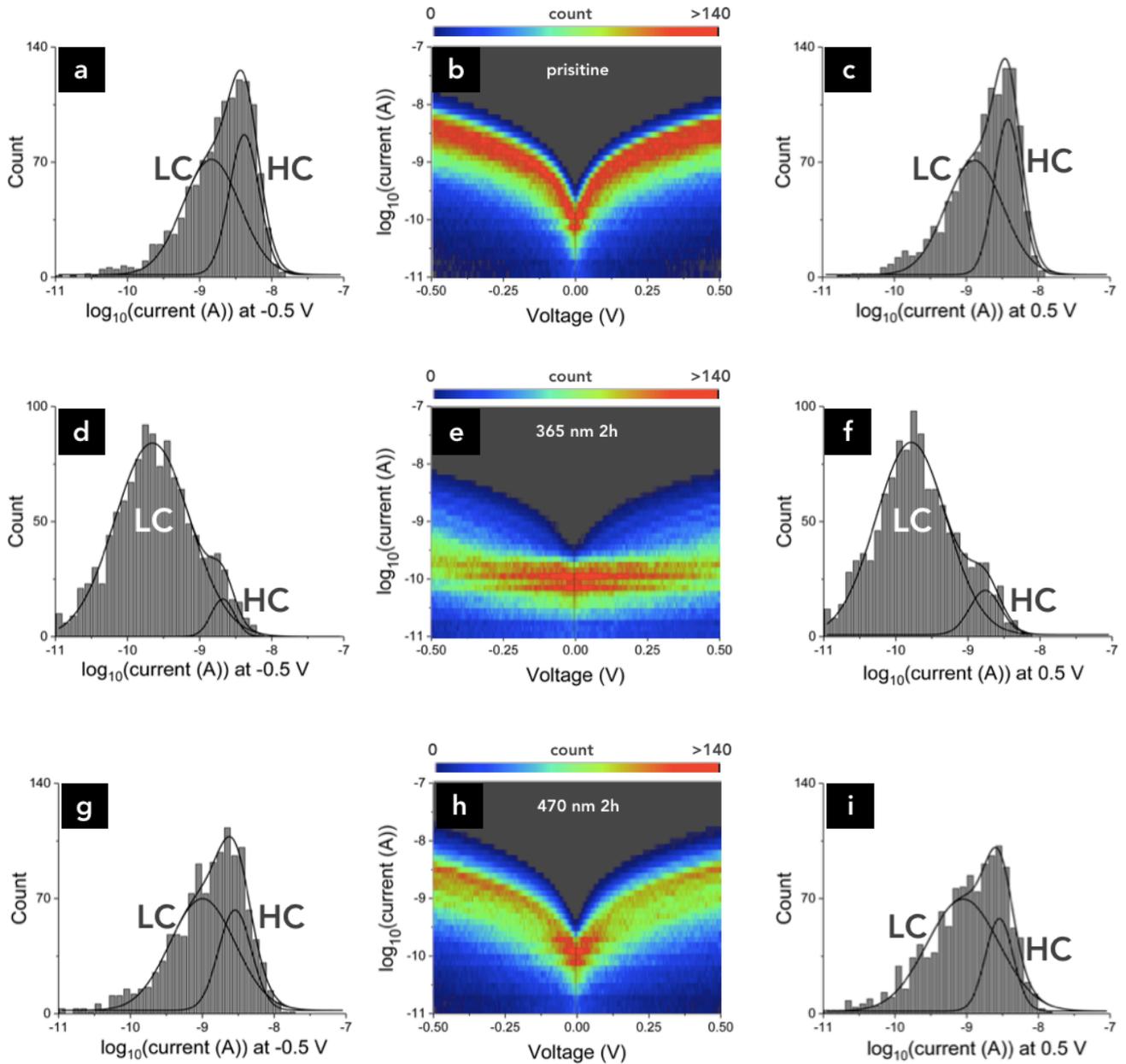

*Figure 3. (a, c)* Histograms and log-normal fits of log(current) at -0.5 V and +0.5 V. *(b)* 2D-histogram of log(current) vs. voltage for Co-AzBT in the pristine state. *(d-f)* Same as (a-c) after 2h of UV irradiation. Note that the horizontal red line between ca. -0.3 and 0.3V corresponds to the preamplifier sensitivity limit of the current-voltage preamplifier (see supporting information). *(g-i)* Same as (a-c) after 2h of blue light irradiation. 1200 I-V traces are used in each histogram.

After 2 hours of UV irradiation (see Methods) to induce the *trans*-AzBT to *cis*-AzBT isomerization, two peaks are still present in the statistical distribution (Fig. 3d-f). We observed two main features; i) the global distribution is shifted to lower current values. At -0.5 and +0.5 V, the LC peaks are now located in the 100 pA range while the HC peaks are slightly shifted and remain in the nA range (HC peak: log-$\mu$ = -8.68 (2.09 × $10^{-9}$ A), log-$\sigma$ = 0.16 at -0.5 V, log-$\mu$ = -8.76 (1.74 × $10^{-9}$ A), log-$\sigma$ = 0.22 at +0.5 V, LC peak: log-$\mu$ = -9.65 (2.24 × $10^{-10}$ A), log-$\sigma$ = 0.52 at -0.5 V, log-$\mu$ = -9.79 (1.62 × $10^{-10}$ A), log-$\sigma$ = 0.49 at +0.5 V). The centers of the HC and LC peaks are now separated by about a decade; and ii) the asymmetry of the peak amplitudes is reversed. At both bias polarities, the LC peak amplitude is higher than that of the HC peak, with a relative weight of 6-9% for the HC peak and 91-94% for the LC peak (Table 1). Thus, after the UV irradiation, the Co-AzBT/PtIr junctions have globally shifted to a lower conductance state. Irrespective to the switching mechanism and the molecule conformation associated to the conductance states (see Discussion section), we can quantify the overall conductance switching ratio of the Co-AzBT/PtIr junctions as the ratio of the mean currents of the most intense, dominant, peaks in the pristine state (i.e., HC peak) and after UV irradiation (i.e., LC peak). The highest ON/OFF conductance ratios are 19 and 23 at -0.5 and +0.5 V, respectively.

|  | V = -0.5 V | | V = +0.5 V | |
| --- | --- | --- | --- | --- |
|  | **HC peak** | **LC peak** | **HC peak** | **LC peak** |
|  | Pristine | | | |
| Area (relative contribution) | 44.02 (39%) | 69.77 (61%) | 44.95 (39%) | 68.90 (61%) |
| log-μ (mean current (A)) | -8.38 (4.17 × $10^{-9}$) | -8.84 (1.45 × $10^{-9}$) | -8.42 (3.80 × $10^{-9}$) | -8.89 (1.29 × $10^{-9}$) |
| log-σ | 0.21 | 0.40 | 0.19 | 0.39 |
|  | 365 nm 2h | | | |
| Area (relative contribution) | 6.56 (6%) | 109.47 (94%) | 10.45 (9%) | 102.23 (91%) |
| log-μ (mean current (A)) | -8.68 (2.09 × $10^{-9}$) | -9.65 (2.24 × $10^{-10}$) | -8.76 (1.74 × $10^{-9}$) | -9.79 (1.62 × $10^{-10}$) |
| log-σ | 0.16 | 0.52 | 0.22 | 0.49 |
|  | 470 nm 2h | | | |
| Area (relative contribution) | 34.84 (31%) | 76.95 (69%) | 27.10 (24%) | 85.65 (76%) |
| log-μ (mean current (A)) | -8.54 (2.88 × $10^{-9}$) | -9.00 ($10^{-9}$) | -8.54 (2.88 × $10^{-9}$) | -9.05 (8.91 × $10^{-10}$) |
| log-σ | 0.23 | 0.45 | 0.19 | 0.50 |

*Table 1.* *Fitted parameters (log-normal) on the histograms of Fig. 3 for the LC (low conductance) and HC (high conductance) peaks. Area under the peaks (a.u.) and relative weight of the peaks (%), log-mean current (log-μ) and corresponding mean current (A), and the log-standard deviation (log-σ).*

This switching is reversible. After blue light irradiation (see Methods) to induce a *cis*-AzBT to *trans*-AzBT isomerization, both the 2D I-V histograms and the 1D histograms at +/- 0.5 V (Fig. 3g-i) are almost similar as the pristine case. As in the pristine state, at both bias polarities the position of the LC and HC peaks are in the nA range with a similar shift (about a half-decade) between them (HC peak: log-μ = -8.54 (2.88 × $10^{-9}$ A), log-σ = 0.23 at -0.5 V, log-μ = -8.54 (2.88 × $10^{-9}$ A), log-σ = 0.19 at +0.5 V; LC peak: log-μ = -9.00 ($10^{-9}$ A), log-σ = 0.45 at -0.5 V, log-μ = -9.05 (8.91 × $10^{-10}$ A), log-σ = 0.50 at +0.5 V). The switch back is probably not complete under this light irradiation condition since the HC/LC relative contributions are 24-31% for the HC peak and 69-76% for the LC peak, *i.e.* a HC peak amplitude slightly lower than for the pristine state (Table 1). Finally, given the estimated number of molecules ( ~ 10, see above), we crudely infer a single molecule conductance of $4.5 \times 10^{-6} G_0 / 1.1 \times 10^{-5} G_0$ (min/max) for the HC peak and $4.1 \times 10^{-7} G_0 / 3.7 \times 10^{-6} G_0$ (min/max) for the LC peak (these values are calculated from the min and max of the mean values reported in Table1, $G_0$ is the quantum of conductance $7.75 \times 10^{-5}$ S). We note that these estimates do not take into account intermolecular interactions in the junctions which are known to significantly influence the conductance of an ensemble of molecules in parallel.[36, 37] We also note that the currents measured for these Co-AzBT/PtIr junctions are lower than for a direct Co/PtIr contact (a factor of at least ~ $10^3$, see Fig. S1 in the supporting information), i.e. there is no short through the SAMs albeit the low molecular packing in the monolayers.

**Theoretical results.**

We consider two Co-AzBT interfacial geometries for the *trans* and *cis* states of AzBT (Scheme 2). For the *trans* state, we calculate the electronic and transport properties of two distinct conformations where the azobenzene moiety is rotated by 180° (*trans1* and *trans2*) with respect to the bithiophene unit (syn- versus anti-orientation). For the *cis* state, we consider two extreme cases where the azobenzene is folded upward (*cis1*) or downward (*cis2*). The molecules are tilted with respect to the substrate to match the SAM thickness deduced experimentally for the *trans* and *cis* states, corresponding to an area per molecule of 180 Å$^2$ (see supporting information).

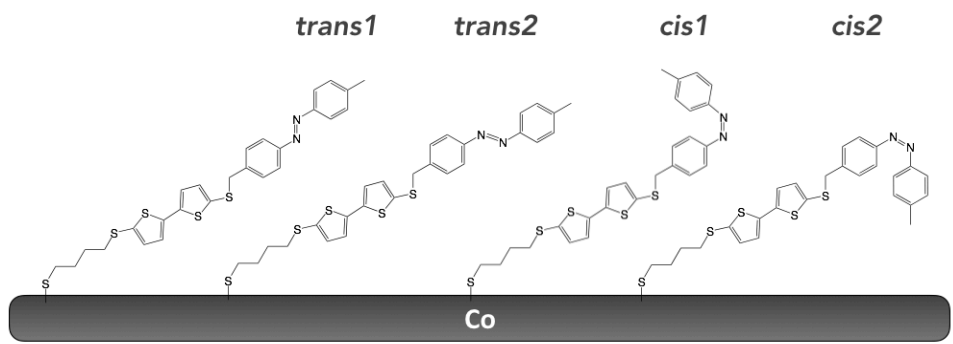

***Scheme 2.*** *Scheme of the four Co-AzBT conformations used for the calculations.*

***Co-AzBT self-assembled monolayers.*** Table 2 collects the work function shift computed for all AzBT conformations together with the values of the molecular contribution $\Delta\Phi_{SAM}$ and the bond dipole $\Delta\Phi_{BD}$ (see Methods and supporting information). The work function of the cobalt (111) surface modified with the compounds in *trans1*, *trans2*, *cis1* and *cis2* conformations are 4.14 eV, 4.39 eV, 3.85 eV and 4.29 eV, respectively, which represents a decrease by -0.93 eV, -0.68 eV, -1.22 eV and -0.78 eV with respect to the bare cobalt (111) surface (calculated to be 5.07 eV in good agreement with the reported experimental value of 5.00 eV[38]). Thus, the *cis1* (*trans2*) form exhibits the higher (lower) work function shift. In order to elucidate the origin of this variation, we compute the two components of the work function shift, namely the molecular contribution ($\Delta\Phi_{SAM}$) and the bond dipole ($\Delta\Phi_{BD}$), as described in the methodology section and supporting information. The results indicate that *cis1* displays the higher molecular contribution of -0.74 eV and a bond dipole of -0.48 eV. A similar value of the bond dipole (-0.47 eV) is obtained for the *trans2* conformation, which promotes, however, the smaller work function shift. The similar amplitude of the bond dipole is explained by the fact that the isomerization of the azobenzene part weakly perturbs the interfacial contact geometry; similar values of BD are also obtained for the two other conformations. Thereby, the shift in the work function is dominated by the difference in the dipole moment between the AzBT forms rather than by the electronic reorganization upon formation of the Co-S bond. This is also evidenced by the magnitude of the corresponding molecular dipole moment calculated according to the Helmholtz equation (Eq. S2 in the Supporting Information). In fact, when going from *trans2* (the form with the smaller work function shift) to *cis1* (the form with the higher work function shift), the intrinsic dipole moment $\mu_{SAM}$ evolves from 1.00 D to 3.52 D, which corresponds to a significant increase of 2.67 D, whereas the dipole moment arising from the bond dipole, $\mu_{BD}$, shows little variation (2.24 D and 2.28 D for *trans2* and *cis1*, respectively).

|            | $\Phi$ (eV) | $\Delta\Phi$ (eV) | $\Delta\Phi_{SAM}$ (eV) | $\Delta\Phi_{BD}$ (eV) | $\mu_{SAM}$ (D) | $\mu_{BD}$ (D) |
|---|---|---|---|---|---|---|
| Co-trans1 | 4.14 | -0.93 | -0.36 | -0.57 | 1.71 | 2.72 |
| Co-trans2 | 4.39 | -0.68 | -0.21 | -0.47 | 1.00 | 2.24 |
| Co-cis1   | 3.85 | -1.22 | -0.74 | -0.48 | 3.52 | 2.28 |
| Co-cis2   | 4.29 | -0.78 | -0.18 | -0.60 | 0.85 | 2.86 |

*Table 2. Calculated work function of Co-SAMs ($\Phi$), work function shift ($\Delta\Phi$) upon SAM deposition, molecular contribution ($\Delta\Phi_{SAM}$), bond dipole ($\Delta\Phi_{BD}$), normal component of the dipole moment ($\mu_{SAM}$, $\mu_{BD}$) obtained from the Helmholtz formula.*

Besides the work function shift, the alignment of the frontier molecular orbitals with respect to the energy of the Fermi level ($E_F$) is another relevant interfacial parameter. Since the HOMO and LUMO levels of the molecules are broadened and mixed with the continuum of states of the cobalt surface, we compute (see Methods) their energy in the SAM by projecting the density of states on the molecules, as displayed in figure 4. The HOMO and LUMO levels are identified as the maximum of the broadened peaks. For all AzBT forms, the HOMO level is closer to the Fermi level than the LUMO, implying that the HOMO level will provide the main transport channel for charge flow through the molecular junctions (*vide infra*).

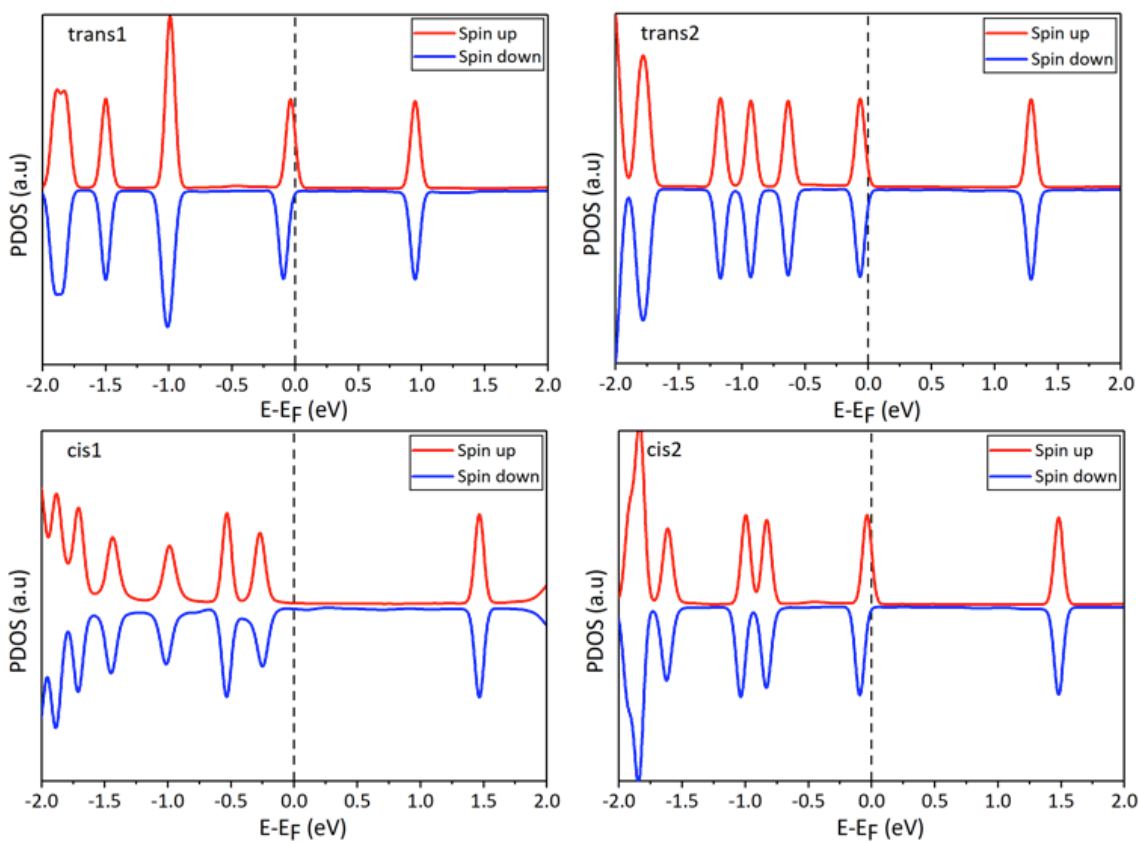

***Figure 4***. *Projected Density of States (PDoS) on the AzBT molecules. The zero energy is set to the Fermi energy.*

Interestingly, *trans1*, *trans2* and *cis2* exhibit quite similar HOMO energies with respect to the Fermi level for spin up (spin down) of -0.03 eV, -0.07 eV and -0.03 eV (-0.09 eV, -0.07 eV and -0.09 eV) respectively, whereas *cis1* displays a HOMO level at lower energy (-0.27 eV for both spin up and spin down). Consistently, the increase in the energy barrier for hole injection (defined here as $E_F-E_{HOMO}$) and the larger work function modification for *cis1* will largely impact the conductance of the corresponding molecular junction compared to the other configurations (*vide infra*).

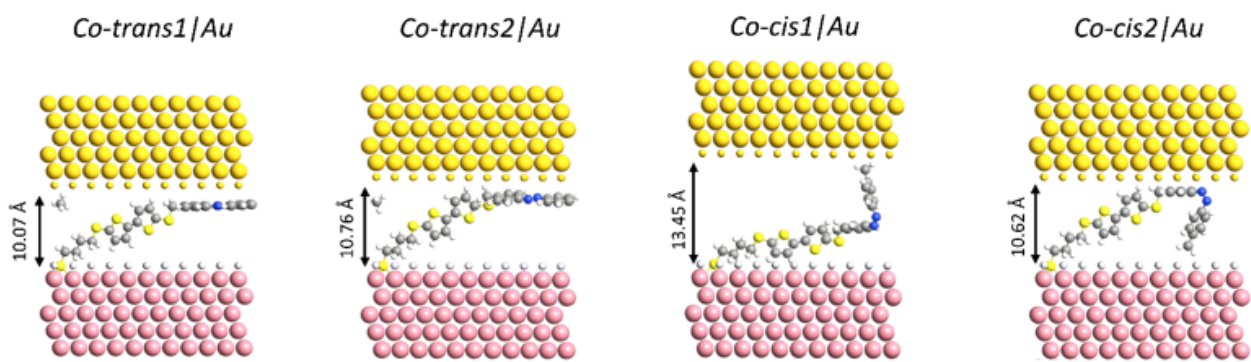

*Figure 5. Optimized AzBT junctions. The calculated junction thickness is also marked. The cis1 form displays the larger thickness.*

***Zero-bias transmission.*** The zero-bias transmission spectra (calculated as the sum of spin-up and spin-down transmission spectra) of the optimized Co-AzBT/Au junctions (Fig. 5) are calculated and plotted in a linear scale in figure 6a (see Methods). Note that the choice of Au top electrodes instead of a typical cobalt electrode is rationalized by two facts: (1) the measurements were performed with a PtIr tip, which is not a ferromagnetic material; and (2) the work function calculated for this top electrode is 4.81 eV, in good agreement with the experimental PtIr work function of 4.86 eV.[39] Many transmission peaks are located below the Fermi level within the energy range from -1.50 to 0 eV while there is only one sharp transmission peak above the Fermi level located at higher energy (about 1.68 eV). This reveals that occupied levels are responsible for the significant transmission peaks involved as possible channels for charge transport through the molecular junctions. For further explicit comparison, we plot in figure 6b the transmission spectra of the Co-AzBT/Au junctions in a log scale within the energy range between -0.50 and 0.50 eV. Although *cis1* exhibits a larger transmission band, the latter lies deeper with respect to the Fermi level (-0.35 eV) whereas the two *trans* and the *cis2* conformations display quite similar transmission coefficients and peak energy positions.

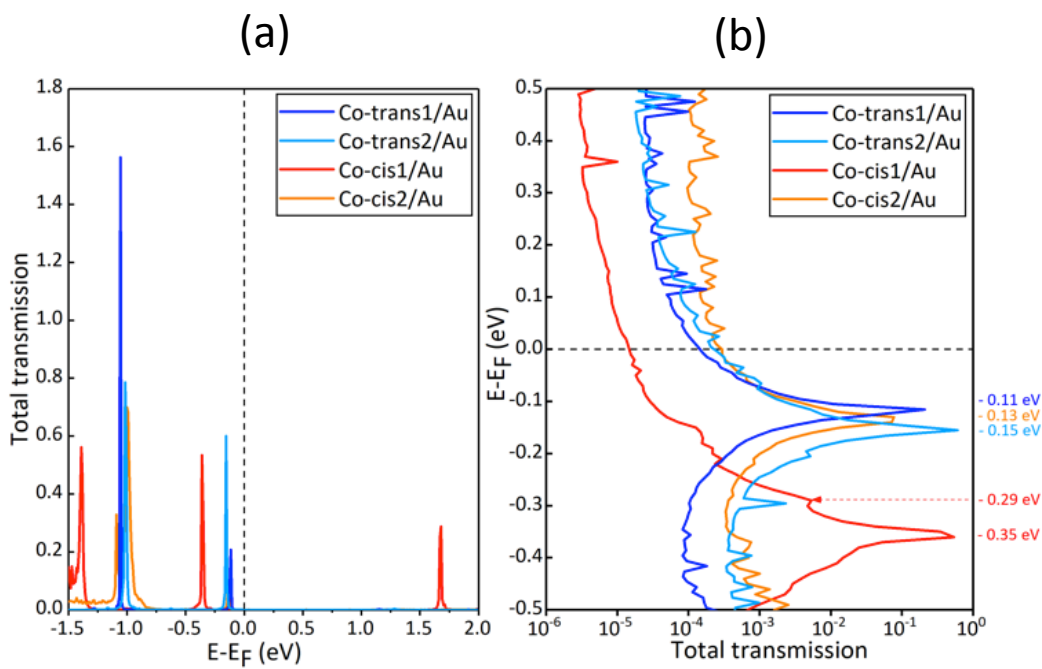

***Figure 6. (a)*** *Linear and **(b)** log scale plot of the transmission spectra at zero bias (sum of spin-up and spin-down transmission spectra) for all AzBT junctions. The dashed line refers to the Fermi energy set to zero.*

To reveal the origin of the peaks in the transmission spectra, the Local Device Density of States (LDDoS), projected onto the real-space transport direction, is calculated at each transmission peak energy (see Methods). As illustrated in figure 7, we clearly recognize the signature of the HOMO shape characteristic of the isolated molecule for the dominant transmission peak of *trans1*, *trans2* and *cis2* located at -0.11 eV, -0.15 eV, -0.13 eV, respectively. In contrast, the LDDoS of the *cis1* junction shows that the sharp peak at -0.35 eV matches the HOMO-1 shape of the isolated molecule, rather than the HOMO level, which is located at a slightly higher energy (-0.29 eV) with a lower transmission coefficient. This can be explained by the increase in the junction thickness and hence the tunneling barrier length in *cis1* that decouples the HOMO state from the top electrode (as indicated by a red arrow in Fig. 7) and thereby makes the HOMO-1 level, mainly localized on the right side of the molecule and filling most of the width of the junction, as the source of the dominant transmission channel (Fig. 7). Overall, the lower transmission coefficient at the Fermi energy of the *cis1* conformation can thus be attributed to an increase in both the injection barrier ($E_F - E_{HOMO}$) and the tunneling barrier length.

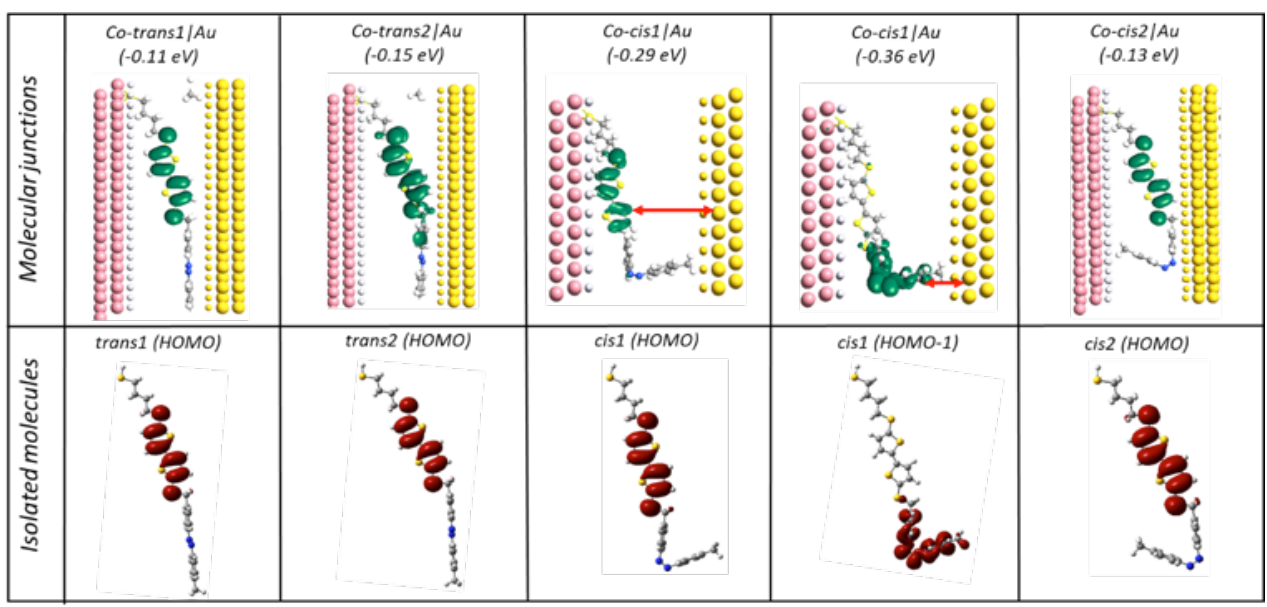

*Figure 7.* Shape of the transmitting occupied orbitals within the junction (LDDoS analysis) and in gas phase (MO analysis). The LDDoS energies (at the peaks of T(E) in Fig. 6) are given with respect to the Fermi level energy set to 0 eV.

***Junctions under bias.*** The I-V curves of the Co-AzBT/Au junctions within the bias window from 0 to 0.5 V were obtained from the Landauer-Büttiker formula[27] by integration of the transmission spectra at equilibrium within the bias window (defined by the magnitude of the applied bias) by considering three extreme scenarios (see Methods and supporting information). The first two scenarios consider that the voltage drop exclusively occurs at the weakest contact (i.e., mechanical C-AFM tip contact) and that the Fermi level of gold top electrode is lowered by -e|V| for the positive voltage range and raised by e|V| in the negative voltage range. In the third scenario, the applied potential is symmetrically distributed, one Fermi level is raised by e|V|/2 and the other lowered by -e|V|/2. The results in Fig.S5 (supporting information) show that the first and second scenario should give rise to a significant rectification of the I-V curve, which is not seen experimentally. We thus conclude that the third scenario is more reasonable and is therefore used to calculate the IV characteristics of the Co-AzBT/Au junctions. As shown in Fig. 8, the current for *cis1* is distinctively lower than for *trans* and *cis2* conformations. The I-V curves of the *trans* isomers (*trans1* and *trans2*) are quite similar, which indicates that the transport properties of these two conformations are indiscernible.

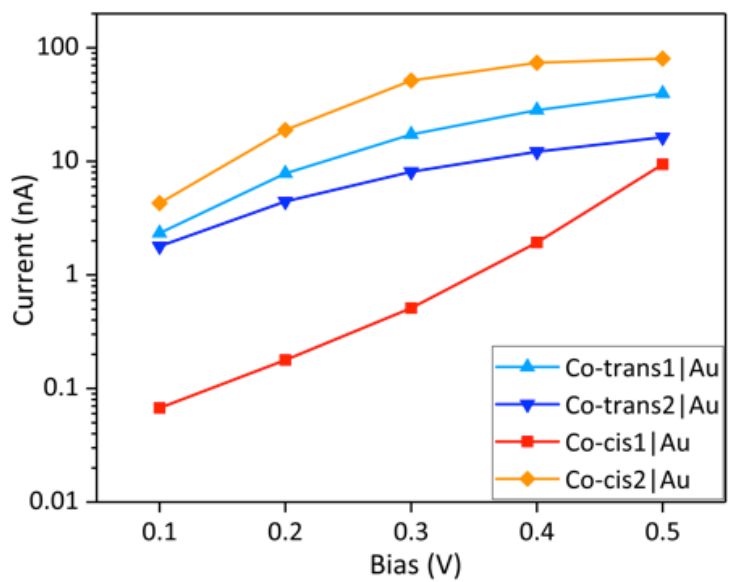

*Figure 8. Semilog scale plot of current (nA) versus bias (V) for the Co-AzBT/Au junctions.*

Altogether, the simulation of Co-AzBT/Au junctions identifies two distinct states (*trans* and *cis*) with different conductance. The ON-state is associated with the *trans* isomers while the OFF-state is attributed to the signature of the *cis1* isomer. To quantitatively evaluate the switching effects between these two forms, we defined an ON/OFF ratio as I (*trans*) /I (*cis1*), where I (*trans*) and I(*cis1*) represent the current through the *trans1* or *trans2* and *cis1* forms, respectively. We stress that the higher current slope that appears beyond 0.3V for *cis1* (see Fig. 8) arises from the integration of the equilibrium transmission spectrum, which is an approximation since any shift of the electronic levels under bias is neglected. Since this feature is not seen in the experimental curve, we speculate that the equilibrium transmission is not accurate beyond 0.3V and hence that our theoretical ratios are more accurate below 0.3V. As displayed in Table 3, the ON/OFF ratios in this low voltage are within the range of 25-44, which is in good agreement with the experimentally measured ON/OFF conductance ratio of about 23 at 0.5 V (Fig. 3 and Table 1).

|  | 0.1V | 0.2 V | 0.3 V | 0.4 V | 0.5 V |
| --- | --- | --- | --- | --- | --- |
| Trans1/cis1 | 34.7 | 44.1 | 33.8 | 14.7 | 4.2 |
| Trans2/cis1 | 26.6 | 24.9 | 15.9 | 6.3 | 1.7 |

*Table 3. The calculated ON/OFF ratio between trans and cis1 forms.*

## DISCUSSION

We clearly report the reversible, albeit partial, conductance switching of AzBT SAMs on Co. For the pristine SAMs, we observe two peaks with a larger contribution of the HC peak. After UV irradiation to induce the *trans*-to-*cis* isomerization, we observe a dominant contribution of the LC peak and a strong decrease in the amplitude of the HC peak. Upon blue light irradiation, the initial situation (*i.e.* the pristine case) is recovered. By comparison with the calculations, we conclude that the LC peak is mainly due to AzBT in the *cis1* conformation, which gives the lower transmission coefficient at $E_F$ (Fig. 6) and the lower conductance (Fig. 8). The HC peak is attributed to the AzBT in the *trans* conformation, the two "model" cases (*trans1* and *trans2*) showing similar properties, which are not experimentally discernible according to theory (see Fig. 8) and given the broad current distribution of the HC peak (Fig. 3). Moreover, it is likely that with the low surface coverage SAM, the organizational disorder favors the coexistence of several conformations. A good agreement is obtained between the experimental and calculated ON/OFF ratios.

This peak identification call for two remarks: i) For the Co-AzBT SAM, we have the conductance $G_{cis} < G_{trans}$. This feature is opposite to our previous observation for the same AzBT molecules on gold ($G_{cis} > G_{trans}$).[20] This is explained by the difference in the SAM organization due to lower density of the SAM on cobalt. On Au surfaces, the AzBT molecules in the *trans* isomer are standing almost upright normal to the surface with the terminal $CH_3$ acting as a tunnel barrier with the CAFM tip, while for the *cis* isomer, the CAFM tip is directly contacting the N=N bond of the azobenzene moiety, leading to a higher conductance $G_{cis}$.[20] In addition, we note that upon *trans*-to-*cis* isomerization, the AzBT SAM thickness decreases on Au, while it increases in the present case, thus modulating the tunnel current accordingly. Several groups have reported $G_{cis} > G_{trans}$ in molecular junctions including various azobenzene derivatives and electrodes (Au, graphene),[40-46] though the opposite situation, $G_{cis} < G_{trans}$, was also reported.[47-50] This discrepancy is not surprising since the geometry of the junction at the atomic level is an important parameter. For example, the two cases can be theoretically explained whether the molecule is chemisorbed at the two electrodes ($G_{cis} < G_{trans}$) or only at one electrode and physisorbed at the other one ($G_{cis} > G_{trans}$).[51] ii) The coexistence of the *trans* and *cis* isomers in the pristine SAM is consistent with the composition in solution. From UV-vis spectroscopy, we have previously observed two isosbestic points (at 297 nm and 380 nm) related to the coexistence of the two isomers in equilibrium.[18] Similarly, our previous study of AzBT SAMs on Au electrodes showed a mixed population (about 55% *cis* and 45% *trans*) in the pristine

SAMs,[20] consistent with the relative weights of the LC (*cis* isomer) and HC (*trans* isomer) peaks for pristine SAMs on Co (Table 1).

After UV light irradiation, in addition to the change in their respective amplitudes, the two LC and HC peaks are also globally shifted to lower currents (by a factor 3-8). This feature has to be linked to the fact that the population of *cis1* isomers increases in the SAMs. Thus, increasing the population of the *cis1* isomers in the SAM does increase the average hole barrier injection ($E_F$-$E_{HOMO}$) at the Co-AzBT interface, as confirmed by the calculations. In addition, the increased average SAM thickness globally reduces the current in the molecular junctions. This average modification of the global electrostatic landscape and layer thickness induced by changes in the population of the *cis* and *trans* isomers in the SAMs (Table 1) appears to be responsible for the global shift of the measured current histograms (Fig. 3). Finally, the *cis2* isomer has electronic properties not discernible from those of the *trans* isomers. However, this conformation is less probable because of the steric hindrance induced by the adjacent molecules that should prevent such a downward folding of the azobenzene moiety. Moreover, the measurements clearly indicate a significant modification of the conductance upon light irradiation (Fig. 3). Accordingly, *cis2* has not been considered in our analysis.

## CONCLUSION

The functionalization of the cobalt surface with photochromic AzBT molecules is structurally characterized and the measurements show the presence of a monolayer on the surface. The light-induced isomerization of the molecules in the SAM induces clearly a reversible change in the conductance histograms of the junctions with a ratio of about 23 between the currents at the maximum of the distributions. By a careful comparison with first-principle calculations, we identify the more conducting ON state to the *trans* isomer, while the *cis* isomer, upward folded, is responsible for the less conducting OFF state. The calculated ON/OFF ratio is in good agreement with the experiments. Albeit not gigantic, this ON/OFF ratio has a reasonable amplitude to make the present Co-AzBT molecular junction a suitable test-bed to assess, at the nanoscale using an UHV CAFM equipped with a magnetic field, the relationship between the spin-polarized electron transport, the molecule conformation in the junction and the molecule-Co interface configurations, the molecule orientations at a ferromagnetic electrode being an important factor of the spin-polarized electron transport.[52]

## METHODS

**Sample fabrication.**

*General conditions of the process.* To avoid the oxidation of the cobalt substrates, the fabrication of the samples (*i.e.* cobalt deposition, preparation of the solutions, grafting of the SAMs and cleaning of the samples) was carried out under nitrogen atmosphere inside a glovebox (MBRAUN) ($H_2O$ and $O_2$ below 5 ppm). The glassware was oven dried at 120°C overnight before immediate insertion inside the glovebox to remove residual adsorbed water on the surfaces. The different steps were quickly sequenced to avoid downtime.

*Cobalt substrate fabrication.* We prepared cobalt substrates by evaporating about 40 nm of cobalt on cleaved (12 ×10 mm) highly phosphorus-doped n-Si(100) substrates, resistivity < $5.10^{-3}$ Ω.cm (from Siltronix), covered by native oxide, cleaned by 5 min sonication in acetone and isopropanol, then rinsed with isopropanol and dried under $N_2$ flow. The evaporation of 99.99% purity cobalt pellets (Neyco) was realized by Joule effect in a vacuum evaporation system (Edwards Auto306) placed inside the glovebox. The cobalt deposition was realized under a $10^{-6}$ mbar vacuum and at a low rate deposition between 2 and 5 Å/s in order to minimize roughness.

***Formation of SAMs on cobalt***. The AzBT molecule was synthesized following a procedure described elsewhere.[18] AzBT SAMs were formed in anhydrous toluene (99.8% from Sigma Aldrich). The solvent was stored over 4 Å molecular sieves (> 5 days), freshly activated (18 h at 250 °C) and degassed with nitrogen (30 min). Freshly prepared cobalt substrates were immediately immersed in a degassed millimolar solution of AzBT in anhydrous toluene for 22 h, in the dark, subsequently rinsed with degassed anhydrous toluene and dried under $N_2$ stream. Note that different samples, all prepared following the same procedure, were used in every experiment carried out in this work, *i.e.* AFM nano-etching, ellipsometry, XPS and UHV CAFM I-V spectroscopy.

***Transfer under controlled atmosphere***. To transfer the samples from the glovebox to the CAFM and the X-ray Photoelectron Spectroscopy (XPS) instrument under UHV, the samples were attached on holders for scanning probe microscopy (Scienta Omicron) by a metallic clamp to form an electrical contact on the cobalt layer. The samples were transferred from the glovebox to the UHV instruments in a homemade hermetic transport container under overpressure of $N_2$ (atmosphere of the glovebox).

**XPS analysis.**

XPS experiments were performed to analyze the chemical composition of the SAMs and to check the residual oxidation state of the cobalt surface. We used a Physical Electronics 5600 spectrometer fitted in an UHV chamber with a residual pressure of $3 \times 10^{-10}$ mbar. The measurements were done using standard procedures (see details in the supporting information).

**Spectroscopic ellipsometry.**

The ellipsometry was used to determine the thickness of the Co-AzBT SAM. To avoid the oxidation of the cobalt in contact with the atmosphere, the sample was placed in a sealed cell (HORIBA) filled with the $N_2$ atmosphere of the glovebox. This container was transferred to an UVISEL ellipsometer (HORIBA) used to record spectroscopic ellipsometry data (see details the supporting information).

**UHV CAFM measurements.**

The CAFM experiments on SAMs under UHV (pressure $10^{-11} - 10^{-9}$ mbar) were performed at room temperature using a VT-SPM microscope (Scienta Omicron). CAFM imaging and local current-voltage (I-V) spectroscopy were carried out using Platinum-Iridium coated probes (SCM-PIC-V2 Bruker), the bias (V) was applied on the substrate. Typically, up to 1200 I-V traces were recorded at a load force of ~ 20 nN to construct 2D I-V histograms (400 I-V recorded on a 20 x 20 grid, pitch = 25 nm, repeated on 3 zones on the SAM, see details in the supporting information). We also used AFM nano-etching to estimate the SAM thickness. The SAM was intentionally indented applying a high load force (140 nN) during a scan. Then, on an enlarged scan of the same zone, the topographic images revealed the presence of a hole from which we estimated the SAM thickness.

**Irradiation setup of the photochromic Co-AzBT SAMs.**

We used a power LED (M365F1 from Thorlabs) for UV light irradiation. This LED has a wavelength centered at 365 nm and a bandwidth of 7.5 nm. For the blue light irradiation, we used a power LED (M470F1 from Thorlabs) which wavelength is centered at 470 nm with a bandwidth of 25 nm. In the UHV CAFM, irradiation of the sample was performed in the entry lock of the instrument (P = $10^{-6}$ mbar $N_2$). An optical fiber was brought close to the viewport. The measured power density at the location of the sample in this configuration was 7.6 and 6.0 mW/cm² for the UV and blue light, respectively.

**Theoretical methodology.**

To calculate the electronic structure and transport properties of Co-AzBT interfaces, different key steps have been followed, as described below (more details in the supporting information).

*Co-AzBT self-assembled monolayers*. First, the geometries of the isolated AzBT molecules are relaxed with forces below 0.01 eV/Å at the density functional theory (DFT) level using the Perdew-Burke-Ernzerhof (PBE) functional within the generalized gradient approximation (GGA),[53] as implemented in the QuantumATK software (details in the supporting information).[54, 55] The Co (111) surface is modeled by a slab of five layers with 11 × 3 cobalt atoms (see figure S4 in the supporting information). A vacuum region of approximately 30 Å is introduced above the surface and 10 Å below it. The coordinates of the bottom three layers were kept fixed in the bulk lattice geometry while the two top layers were relaxed until the final forces acting on the atoms are less than 0.02 eV/ Å. For this geometric relaxation, we use the Perdew-Burke-Ernzerhof (PBE) functional within the spin generalized gradient approximation (SGGA),[53, 56] a double-zeta plus polarization basis set, a density mesh cutoff of 100 Ha and a (2×8×1) k-sampling. Once the geometry is relaxed, the electronic structure and by extension the work function of the surface are computed using a 6×24×1 k-sampling, a 100 Ha mesh cutoff and a single-zeta plus polarization (SZP) basis set (see supporting information). To build the Co-AzBT SAMs, the relaxed molecules are tilted to fit the measured SAM thickness and then anchored on the cobalt surface through a sulfur atom. The SAM structures are optimized by relaxing the molecules forming the SAMS and the top two metal layers until forces are below 0.04 eV/ Å. The electronic structure and the SAM induced work function shift are then computed following the same method as used for the calculations on the bare cobalt surface, with a DZP basis set for the valence molecular orbitals (see details in the supporting information).

***Co-AzBT molecular junctions***. In a final stage, we add a second electrode on the top side of the molecular layer to build a single heterogeneous molecular junction Co-AzBT/Au. This second electrode is created by conserving the lattice parameters of the cobalt (111) surface and by converting the cobalt atoms to gold atoms to avoid problems related to lattice incommensurability. A van der Waals contact is assumed between the molecular layer and the top electrode, with an interatomic distance determined as the sum of van der Waals radii of the hydrogen and gold atoms (2.86 Å). The electronic transport calculations of the Co-AzBT/Au junctions were performed by the combination of DFT to the Non-Equilibrium Green's Function (NEGF) formalism,[26] as implemented in QuantumATK Q-2019.12-SP1 package (details are reported in the supporting information).[54, 55] This formalism has been widely recognized as a robust tool to rationalize the experimental results and predict new features and trends for charge transport in molecular junctions.[57-59] Finally, the I-V characteristics have been calculated on the basis of Landauer-Büttiker formalism,[27] that links the transmission spectrum to the current in a coherent transport regime (see the supporting information).

## ASSOCIATED CONTENT

**Supporting Information**. This material is available free of charge via the Internet: Additional characterization protocols (XPS, ellipsometry, CAFM) and theoretical methods, XPS spectra for cobalt surface exposed to air and cobalt functionalized with AzBT SAM and tables with the fitting parameters for XPS deconvolution.

## AUTHOR CONTRIBUTIONS

L.T. and D.G. fabricated and characterized the SAMs. L.T. performed the UHV C-AFM measurements. X.W. performed the XPS measurements. I.A. performed the theoretical calculations under the supervision of C.v.D. and J.C. The experimental work done by L.T. is part of his PhD thesis supervised by S.L. and T.M. The project was conceived by D.V. and managed by S.L. and D.V. The manuscript was written by D.V. with the contributions of all the authors. All authors have given approval of the final version of the manuscript.

# These authors (L.T. and I.A.) contributed equally to this work.

## CONFLICT OF INTEREST

The authors declare no competing financial interest.

**ACKNOWLEDGMENTS**

This work has been financially supported by the French National Research Agency (ANR), project SPINFUN ANR-17-CE24-0004 and partly by Renatech. We are grateful to the Philippe Blanchard's group at Moltech-Anjou (CNRS, U. Angers) for the AzBT synthesis. We acknowledge D. Deresmes for his valuable help with the UHV CAFM instrument, J.L. Caudron, E. Galopin, L. Fugère for the XPS measurements, Y. Deblock for ellipsometry, and J.-M. Mallet for the manufacturing of mechanical parts of the transport container. The work of I.A. is supported by the Belgian National Fund for Scientific Research (F.R.S.-FNRS) thanks to the project SPINFUN (Convention T.0054.20). We also acknowledge the Consortium des Équipements de Calcul Intensif (CÉCI) funded by the Belgian National Fund for Scientific Research (F.R.S.-FNRS) for providing the computational resources. J.C. is an FNRS research director.

# Conductance Switching of Azobenzene-Based Self-Assembled Monolayers on Cobalt Probed by UHV Conductive-AFM.

Louis Thomas,[1,#] Imane Arbouch,[2,#] David Guérin,[1] Xavier Wallart,[1] Colin van Dyck,[2] Thierry Mélin,[1] Jérôme Cornil,[2,*] Dominique Vuillaume[1,*] and Stéphane Lenfant[1,*]

*1) Institute of Electronics Microelectronics and Nanotechnology (IEMN), CNRS, University of Lille, Avenue Poincaré, Villeneuve d'Ascq, France.*

*2) Laboratory for Chemistry of Novel Materials, University of Mons, Place du Parc 20, Mons, Belgium.*

# Supporting Information

## 1. XPS characterization.

High resolution XPS spectra were recorded with a monochromatic $Al_{K\alpha}$ X-ray source ($h\upsilon$ = 1486.6 eV), a detection angle of 45° as referenced to the sample surface, an analyzer entrance slit width of 400 µm and with an analyzer pass energy of 12 eV. In these conditions, the overall resolution as measured from the full-width half-maximum (FWHM) of the Ag $3d_{5/2}$ line is 0.55 eV. Background was subtracted by the Shirley method.[1] The peaks were decomposed using Voigt functions and a least squares minimization procedure.

## 2. Ellipsometry.

The system acquired a spectrum ranging from 2 to 4.5 eV (corresponding to 300 to 750 nm) with intervals of 0.1 eV (or 15 nm). To use the cell filled with $N_2$, data were taken at an angle of incidence of 60 ± 1° while the compensator was set at 45°. We fitted the data with DeltaPsi 2 data analysis software by a regression analysis to a film on a substrate model as described by their thickness and their complex refractive indexes. First, we recorded a background before monolayer deposition for a cobalt substrate freshly evaporated on the silicon substrate. We acquired three reference spectra at three different places of the surface spaced of few mm. Secondly, after the monolayer deposition, we acquired once again three spectra at three different places of the surface. We used a 2-layer model (substrate/SAM) to fit the measured data and to determine the SAM thickness. We used the previously measured optical properties of the cobalt substrate (background) and we fixed the refractive index of the organic monolayer at 1.50. The usual values in the literature for the refractive index of organic monolayers are in the range 1.45-1.50.[2, 3] The three spectra measured on the sample were fitted separately using each of the three reference spectra, giving nine values for the SAM thickness. We calculated the mean value from this nine thickness values and the thickness incertitude corresponding to the standard deviation.

### 3. CAFM measurements.

CAFM imaging and local current-voltage (I-V) spectroscopy were carried out using Platinum-Iridium coated probes SCM-PIC-V2 (Bruker), tip radius $R$ = 25 nm, spring constant $k$ = 0.1 N/m. In all our measurements, bias (V) was applied on the substrate and the current was recorded with an external DLPCA-200 amplifier (FEMTO) at a gain of 1μA/V, which set the current sensitivity limit at ca. $10^{-10}$ A. The force applied on the sample by the tip was calculated from approach-retract curves prior to each experiment to ensure the correct vertical deflection set point with respect to the fixed force set point (F). The load force was set at ~ 20 nN for all the *I-V* measurements, a lower value leading to too many contact instabilities during the *I-V* measurements. Albeit larger than the usual load force (2-5 nN) used for CAFM on SAMs, this value is below the limit of about 60-70 nN at which the SAMs start to suffer from severe degradations. For example, a detailed study (Ref. 4) showed a limited strain-induced deformation of the monolayer (≤ 0.3 nm) at this used load force. The same conclusion was confirmed by our own study comparing mechanical and electrical properties of alkylthiol SAMs on flat Au surfaces and tiny Au nanodots.[5] Moreover, we checked by topographic AFM that a load force of 20 nN is not indenting the SAM.[6] Current-voltage spectra (I-V) were acquired on several grids (20 x 20 points spaced by 25 nm from each other), each grid spaced of a few mm. Data were processed using Gwyddion[7] and WSxM[8] software. We typically recorded 1200 I-V traces, used without data selection to construct 2D histogram of the decimal logarithm of the current versus voltage. For a given bias, we fitted the current histogram with log-normal distributions and extracted the relevant statistical quantities: the log-mean current (log-µ) and the log-standard deviation (log-σ). Topographic and current images (at a given bias) were also recorded simultaneously. As usually reported in literature[4, 9-11] the contact radius (a) between the C-AFM tip and the SAM surface, and the SAM elastic deformation (δ) are estimated from a Hertzian model:[12]

$$a^2 = \left( \frac{3RF}{4E^*} \right)^{2/3} \tag{S1}$$

$$\delta = \left( \frac{9}{16R} \right)^{1/3} \left( \frac{F}{E^*} \right)^{2/3} \tag{S2}$$

with F the tip load force (20 nN), R the tip radius (25 nm) and E* the reduced effective Young modulus defined as:

$$E^* = \left( \frac{1}{E^*_{SAM}} + \frac{1}{E^*_{tip}} \right)^{-1} = \left( \frac{1-v^2_{SAM}}{E_{SAM}} + \frac{1-v^2_{tip}}{E_{tip}} \right)^{-1} \tag{S3}$$

In this equation, $E_{SAM/tip}$ and $\nu_{SAM/tip}$ are the Young modulus and the Poisson ratio of the SAM and C-AFM tip, respectively. For the Pt/Ir (90%/10%) tip, we have $E_{tip}$ = 204 GPa and $\nu_{SAM/tip}$ = 0.37 using a rule of mixture with the known material data.[13] These parameters for the AzBT SAM are not known and, in general, they are not easily determined in such a monolayer material. Thus, we consider the value of an effective Young modulus of the SAM $E^*_{SAM}$ = 38 GPa as determined for the "model system" alkylthiol SAMs from a combined mechanics and electron transport study.[4] With these parameters, we estimate a = 2.3 nm (contact area = 16.6 nm²) and δ = 0.2 nm. With a molecular packing density of 1.8 nm²/molecule (see theory, below §6), we infer that about 10 molecules are measured in the Co-AzBT/PtIr junction.

We also checked that there is no short through the SAMs by comparing the I-Vs for the Co-AzBT/PtIr tip (Fig. 3, main text) with the typical I-Vs for a direct contact of the C-AFM tip on Co surface (Fig. S1), which show a current reaching the compliance limit of the preamplifier (here 10 μA) at V < 50 mV.

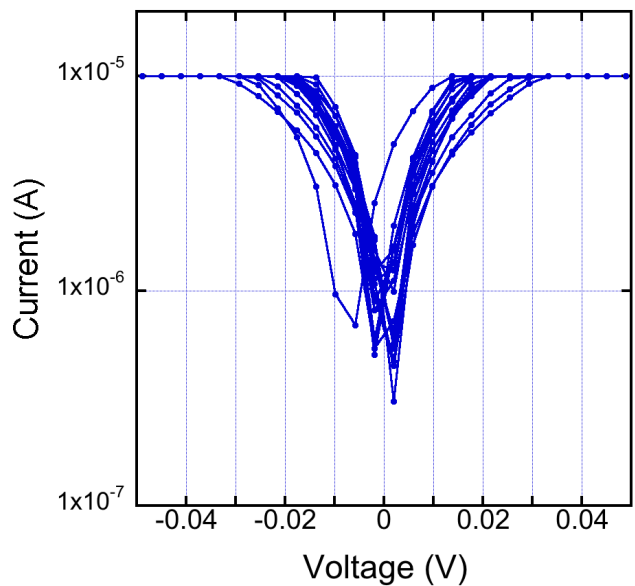

*Figure S1. Typical I-Vs (20 traces) for the direct contact of the PrIr C-AFM tip on a Co surface.*

## 4. XPS analysis of air-exposed cobalt

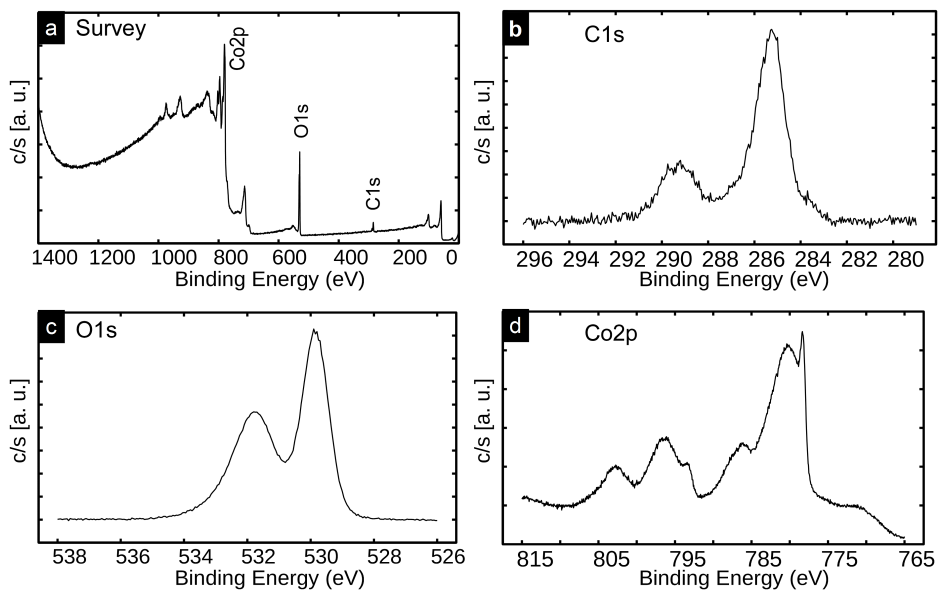

*Figure S2.* XPS spectra of air-exposed cobalt: **(a)** survey, **(b)** C1s, **(c)** O1s, **(d)** Co2p.

| Spectra | Identification | Energy (eV) |
|---|---|---|
| C1s | C-C | 285.2 |
| C1s | C-O-C | 286.3 |
| C1s | C-C=O | 289.3 |
| O1s | CoO, $Co_3O_4$ | 529.8 |
| O1s | C-O, $Co(OH)_2$ | 531.8 |
| Co2p | Co metal 2p3/2 | 778.2 |
| Co2p | CoO, $Co_3O_4$, $Co(OH)_2$ 2p3/2 | 780.3 |
| Co2p | satellite | 785.8 |
| Co2p | Co metal 2p1/2 | 793.4 |
| Co2p | CoO, $Co_3O_4$, $Co(OH)_2$ 2p1/2 | 796.4 |
| Co2p | satellite | 802.6 |

*Table S1.* Identification of the peaks in the XPS analysis of air-exposed cobalt.

*Peak assignation from Refs. 14-18*

## 5. XPS analysis of Co-AzBT SAM

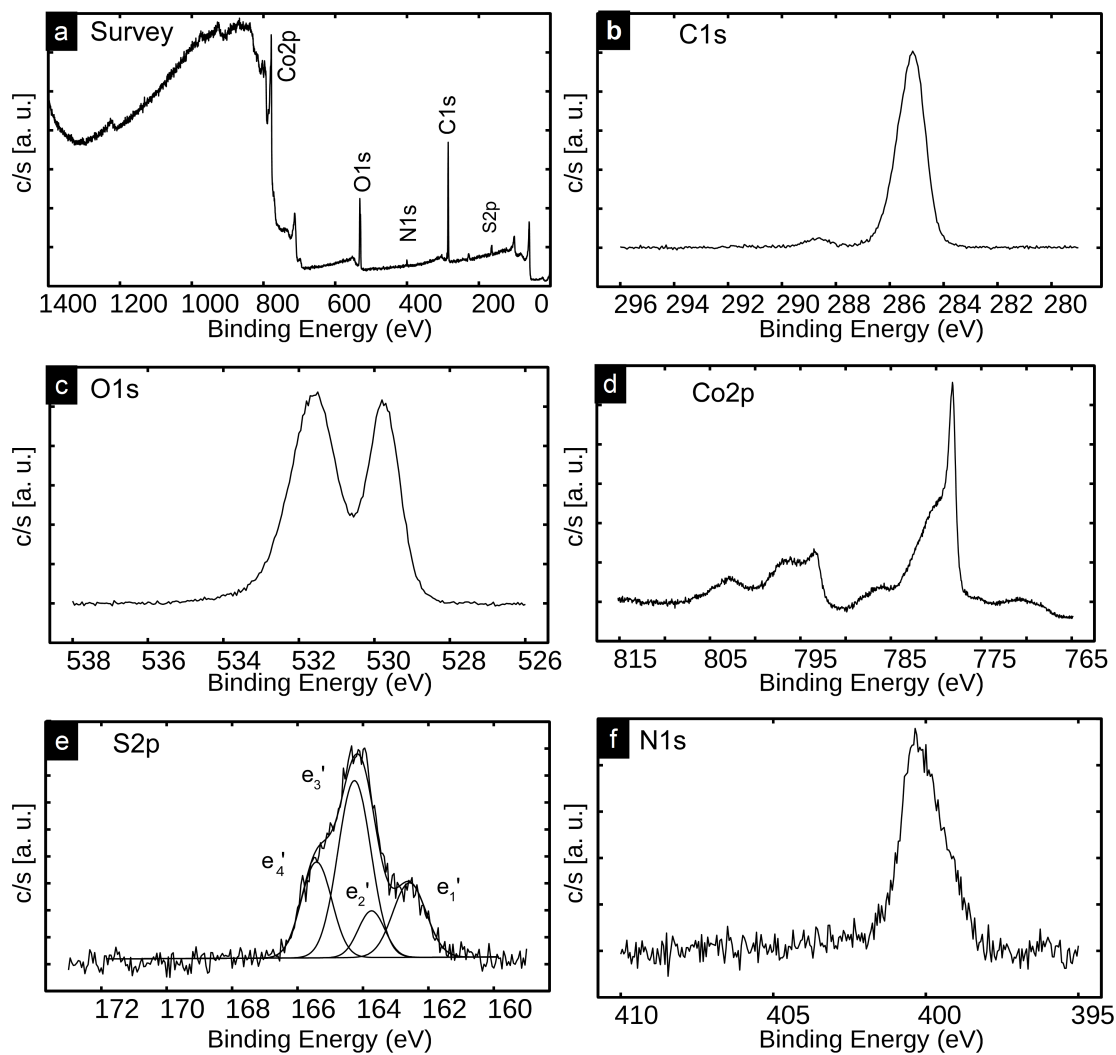

*Figure S3.* XPS spectra of Co-AzBT: **(a)** survey, **(b)** C1s, **(c)** O1s, **(d)** Co2p, **(e)** S2p, **(f)** N1s.

The O1s spectra is composed of two peaks at binding energies of 529.7 eV and 531.6 eV. These values are close to those observed by Chuang et al.[19] on cobalt surface exposed to air (529.5 eV and 530.8 eV). For these authors these peaks are associated to the oxidized cobalt. From the survey spectra (Fig. S3a), a smaller amount of oxygen is detected on Co-AzBT than on the air-exposed sample (Fig. S2a). For the Co-AzBT XPS analysis (Table S2), C1s peaks were observed approximately at the same binding energy as on a cobalt surface exposed to the air without SAM (see Table S1). The first peak associated to C-C and C-S bonds and located at 285.1 eV was also observed on the sample of air-oxidized cobalt without SAM at 285.2 eV. However, the peak area is lower for the air-oxidized cobalt (divided by 4.8) compared to the sample with the SAM (Co-AzBT), with a peak area of 5192 (a.u.) and 1087 (a.u.) for Co-AzBT and air-oxidized cobalt respectively. So, the peak at 285.1 eV observed on Co-AzBT is a contribution of the carbon atoms of the AzBT molecule but also of the surface contamination. If we suppose that the contamination is the same for both the samples, we estimate that the peak observed at 285.1 eV on Co-AzBT is due to 21 % to the contamination and 79 % to the AzBT molecule grafted on the cobalt surface.

| Spectra | Peak | Identification | Energy (eV) |
|---|---|---|---|
| C1s | | C-C, C-S | 285.1 |
| C1s | | C-O-C | 285.6 |
| C1s | | C-C=O | 288.7 |
| O1s | | CoO, $Co_3O_4$ | 529.7 |
| O1s | | C-O, $Co(OH)_2$ | 531.6 |
| Co2p | | Co metal 2p3/2 | 778.3 |
| Co2p | | CoO, $Co_3O_4$ 2p3/2 | 780.5 |
| Co2p | | satellite | 785.1 |
| Co2p | | Co metal 2p1/2 | 793.4 |
| Co2p | | CoO, $Co_3O_4$ 2p1/2 | 796.4 |
| Co2p | | satellite | 802.6 |
| S2p | $e_1'$ | CoS S2p3/2 | 162.6 |
| S2p | $e_2'$ | CoS S2p1/2 | 163.7 |
| S2p | $e_3'$ | CS S2p3/2 | 164.3 |
| S2p | $e_4'$ | CS S2p1/2 | 165.4 |
| N1s | | N in azobenzene | 400.2 |

*Table S2. Identification of the peaks in the XPS analysis of Co-AzBT.*

*Peak assignation from Refs. 14-18*

## 6. Full theoretical methodology.

***Co-AzBT self-assembled monolayers***. First, the geometries of the isolated AzBT molecules are relaxed with forces below 0.01 eV/Å at the density functional theory (DFT) level using the Perdew-Burke-Ernzerhof (PBE) functional within the generalized gradient approximation (GGA),[20] as implemented in the QuantumATK software.[21, 22] The valence electrons are described within the LCAO approximation, with a double-zeta plus polarization basis set (DZP) whilst the core electrons are described by the norm-conserving Troullier-Martins pseudopotentials.[23] We use a density mesh cutoff of 100 Ha and multipole boundary conditions to correct for the interaction between the single molecules introduced in a large unit cell and their images generated by the Periodic Boundary Conditions (PBC). The Co(111) surface is modeled by a slab of five layers with 11 × 3 cobalt atoms in each layer and lattice parameters a= 27.645 Å, b= 7.539 Å and α= 120°. This corresponds to a theoretical area per molecule of 180 Å$^2$ (see Fig. S4). A vacuum region of approximately 30 Å was introduced above the surface and 10 Å below it. The coordinates of the bottom three layers were kept fixed in the bulk lattice geometry while the two top layers were relaxed until the final forces acting on the atoms are less than 0.02 eV/Å. For this relaxation, we use the Perdew-Burke-Ernzerhof (PBE) functional with the spin generalized gradient approximation (SGGA),[20, 24] a double-zeta plus polarization basis set, a density mesh cutoff of 100 Ha and a (2×8×1) k-sampling.

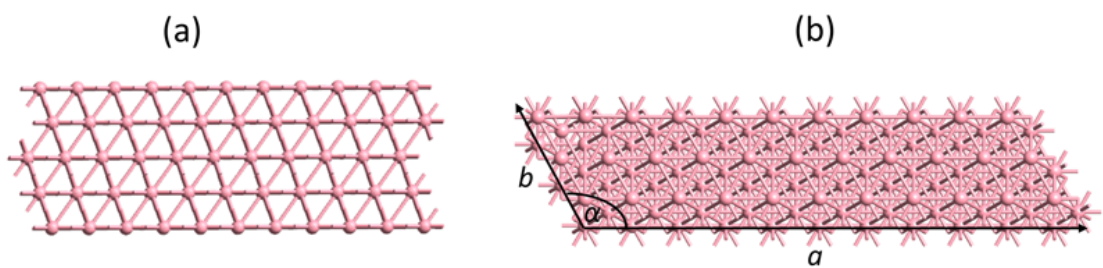

*Figure S4. (a) side and (b) top view of the Co (111) slab.*

Once the geometry is relaxed, the electronic structure and by extension the work function of the surface are computed using a 6×24×1 k-sampling, a 100 Ha mesh cutoff and a single-zeta plus polarization (SZP) basis set. We reduce the basis set here to be consistent with electronic structure computations achieved on larger systems such as SAMs and molecular junctions. For the resolution of Poisson's equation, Neumann (fixed potential gradient) and Dirichlet (fixed potential) were imposed on the left- and right-hand side of the slab, respectively, combined with periodic boundary conditions (PBC) in the in-plane directions. These mixed PBCs provide an alternative to the widely used slab dipole correction.[25, 26] The Fermi energy is calculated initially for cobalt so that its energetic position ensures that the occupation of the density of states of the system equals the total number of electrons. Within this framework, we compute a work function of 3.78 eV for the Co(111) surface, a value too low compared to the experimental value of 5.00 eV.[27] This result is fully expected due to the use of short range localized atomic orbitals that induces an artificial push-back effect at the interface and thus reduces the metal work function. To solve this numerical problem, we added a layer of ghost atoms in the vacuum next to the surface to better describe the tail of the electron density, an approach already implemented in QuantumATK Q-2019.12-SP1.[21] Several tests have been performed in order to identify the nature of the ghost atoms, the converged distance from the top layer, the basis set and the pseudopotential that provide good accuracy at reasonable computational times. The obtained results indicate that the use of a layer of platinum ghost atoms described by medium basis sets and PseudoDojo pseudopotentials[28] at a distance of 1.6 Å from the top cobalt layer provides a work function of 5.07 eV, which is in very good agreement with the reported experimental value of 5.00 eV.[27]

To build the Co-AzBT SAMs, the molecules are tilted to fit the measured SAM thickness and then anchored on the cobalt surface through a sulfur atom. The SAM structures are optimized by relaxing the molecules forming the SAMS and the top two metal layers until forces are below 0.04 eV/Å. The exchange correlation SGGA.PBE functional has been used, with a (DZP) basis set for valence cobalt electrons (i.e., to describe the valence molecular orbitals), a (2×8×1) k-point sampling and a mesh cutoff of 100 Ha. The electronic structure and the SAM induced work function shift are then computed following the same method as used for the calculations on the bare cobalt surface, with a DZP basis set for the valence molecular orbitals. As described for the metal surface, a second ghost layer of gold atoms has been also positioned at -0.1 Å below the top molecular atoms to reduce the artificial push-back effect and prevent any artificial creation of bond dipole due to BSSE (Basis Set

Superposition Errors).[29] These two effects may induce errors larger enough to prevent a direct comparison between the work function shifts of the different systems.

The change in the work function upon SAM deposition, generally referred to as 'vacuum level shift' is well studied in the literature[30, 31] and has two main possible origins: (1) the intrinsic dipole moment of the molecules, reduced by the induced dipole arising from the depolarization effects triggered by the adjacent molecules;[32] and (2) the interfacial electronic reorganization upon formation of the Co-S bond resulting in a bond dipole in the vicinity of the interface. The apparent work function shift is then simply described as the sum of a molecular contribution ($\Delta\Phi_{SAM}$) and a bond dipole ($\Delta\Phi_{BD}$):

$$\Delta\Phi = \Delta\Phi_{SAM} + \Delta\Phi_{BD} \tag{S4}$$

The relative intensities of these two contributions depend on whether we adopt a neutral or a radical approach to describe the bonding of thiol to the metal surface. The neutral approach describes the adsorption as the replacement of a bond (the S-H bond) with a S-Co bond whereas the radical scenario considers the formation of a new covalent bond between the radical and the cobalt surface. For the sake of consistency with our previous studies,[33-35] we adopt here the 'radical scenario' for computing the molecular contribution and the bond dipole. To this end, we compute the electrostatic potential profile across a layer of radicals obtained by removing the cobalt atoms from the full system while keeping the same geometry for the molecular part; we then compute the energy difference between the left and right side of the layer, which corresponds directly to $\Delta\Phi_{SAM}$. The value of BD is obtained by subtracting $\Delta\Phi_{SAM}$ from the total shift. We note for sake of completeness that the amplitude of the work function shift can be related to the corresponding dipole moment by the Helmholtz equation:[36, 37]

$$\Delta\Phi = \mu / \varepsilon_0 A \tag{S5}$$

where µ is the dipole moment in the unit cell, projected along the axis normal to the surface, $\varepsilon_0$ is the vacuum permittivity and A is the area per dipole.

***Co-AzBT molecular junctions***. In a final stage, we add a second electrode on the top side of the molecular layer to build a single heterogeneous molecular junction Co-AzBT/Au. This second electrode is created by conserving the lattice parameters of the cobalt (111) surface and by converting the cobalt atoms to gold atoms to avoid problems related to lattice incommensurability. The choice of this type of electrode instead of a typical cobalt electrode is rationalized by two facts: (1) the measurements were performed with a PtIr tip, which is not a ferromagnetic material; and (2) the work function calculated for this top electrode is 4.81 eV, which is in good agreement with the experimental PtIr work function of 4.86 eV.[38] Note that this work function has been calculated by adding a ghost layer of gold atoms at a distance of 1.7 Å from the top gold layer of the second electrode.

A van der Waals contact is assumed between the molecular layer and the top electrode, with an interatomic distance determined as the sum of van der Waals radii of the hydrogen and gold atoms (2.86 Å). The electronic transport calculations of the Co-AzBT/Au junctions were performed by the combination of DFT to the Non-Equilibrium Green's Function (NEGF) formalism,[39] as implemented in QuantumATK Q-2019.12-SP1 package.[21, 22] This formalism has been widely recognized as a robust tool to rationalize the experimental results and predict new features and trends for charge transport in molecular junctions.[35, 40, 41] The exchange-correlation potential is described with the SGGA.PBE functional[20] whereas the Brillouin zone was sampled with a (3×12×100) k-sampling. We expand the valence electrons in a single zeta plus polarization basis set (SZP) for metal atoms (gold, gold ghost and cobalt), and a double zeta polarization basis set (DZP) for the other atoms. The core electrons are frozen and described by the norm-conserving Troullier−Martins pseudopotentials.[23] As for SAMs calculations, the platinum ghost atoms are described by a medium basis set and PseudoDojo potentials.

[28] The mesh cutoff was set to 80 Ha with a temperature of 300 K. These parameters have been carefully tested to ensure the convergence of the transmission spectrum.

The I-V characteristics have been calculated on the basis of the Landauer-Büttiker formalism,[42] that links the transmission spectrum to the current in a coherent transport regime. When a bias is applied, the current is calculated via the integration of the transmission spectrum within a bias window defined by a Fermi-Dirac statistics in the left and right electrodes:

$$I(V) = \frac{2e}{h} \int T(E) \left[ f(E - \mu_L) - f(E - \mu_R) \right] dE \quad (S6)$$

Where T(E) is the transmission spectrum, E the incident electron energy, $f$ the Fermi-Dirac function $f(E-\mu) = [1 + exp(E-\mu)/kT]^{-1}$, $\mu_{L/R}$ the chemical potential of the left/right electrode, e the elementary charge, $h$ the Planck constant, $k$ the Boltzmann constant and V the applied bias, with $|\mu_L - \mu_R| = eV$.

It is important to note that for an accurate estimation of the current, the transmission spectrum T(E) should be calculated in a self-consistent way for each bias. However, it is possible to obtain a reasonable approximation for the current at low bias by using the transmission spectrum at zero bias. This approximation is required here for our junctions with a large unit cell and a spin-polarized electrode because the self-consistent calculations become very time consuming. Accordingly, the current-voltage properties of Co-AZBT/Au were predicted by using the transmission calculated at equilibrium.

**7. Junctions under bias**

The computed ON/OFF ratios can be affected by the voltage drop profile across the junction. This voltage drop is hard to guess and should be computed theoretically. However, as explained in the methodology, these systems were computationally too heavy to be driven out of equilibrium. To get a qualitative idea of the ratio expected from theory, we thus integrated the equilibrium transmission by considering three extreme scenarios (see Fig. S5):

1. Under bias, one Fermi energy is fixed and the other one is driven up by $e|V|$.

2. Under bias, one Fermi energy is fixed and the other one is driven down by $-e|V|$.

3. Under bias, one Fermi level is driven up by $e|V|/2$ and the second level is driven down by $-e|V|/2$.

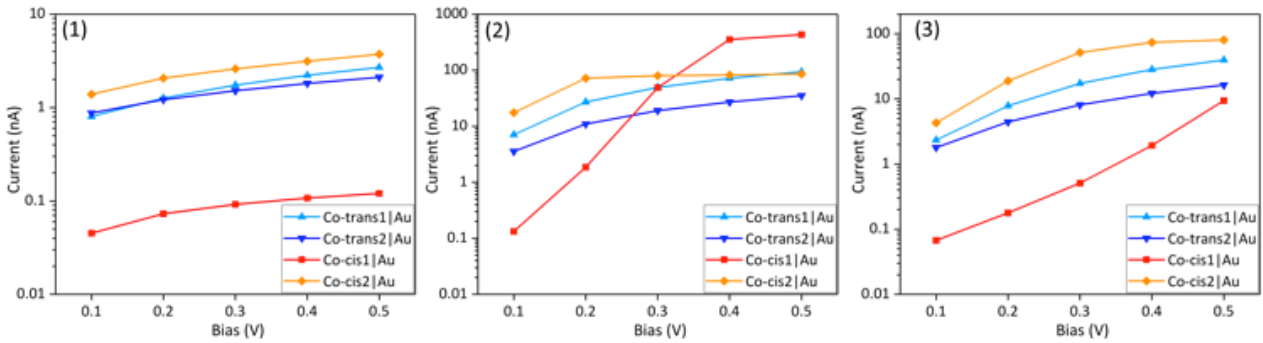

*Figure S5.* *Semilog scale plot of current (nA) versus bias when (1) a single Fermi level is driven up by $e|V|$, (2) a single Fermi level is driven down by $-e|V|$ and (3) one Fermi level is driven up by $e|V|/2$, and the second level is driven down by $-e|V|/2$.*

With our simplified methodology, the first and second scenario should give rise to a significant rectification as shown by the difference of the calculated I-V shapes between cases (1) and (2), because the HOMO and LUMO levels are not located symmetrically with respect to the Fermi energy (Fig. 6, main text), which is not seen experimentally. We thus conclude that the symmetrical bias repartition, scenario (3), is more reasonable. Noteworthy, the higher slope that appears beyond 0.3V for *cis1* (see Fig S5-c) arises from the integration of the equilibrium transmission spectrum, which is clearly an approximation. Since this feature is not seen in the experimental curve, we speculate that the equilibrium transmission is not accurate beyond 0.3V and hence that our theoretical ratios are more accurate below 0.3V. We report the predicted ratios corresponding to the three scenarios in Table S3. In this low voltage range, the three scenarios predict ratios between *trans* and *cis1* between 14 and 53, hence with the same order of magnitude as the experimental values.

| Scenarios | Ratios | 0.1 V | 0.2 V | 0.3 V | 0.4 V | 0.5 V |
|---|---|---|---|---|---|---|
| (a) | Trans1/cis1 | 17.8 | 17.3 | 18.9 | 20.6 | 22.4 |
| | Trans2/cis1 | 19.4 | 16.7 | 16.5 | 16.9 | 17.5 |
| (b) | Trans1/cis1 | 53.0 | 14.6 | 1.0 | 0.2 | 0.2 |
| | Trans2/cis1 | 26.8 | 5.9 | 0.4 | 0.2 | 0.2 |
| (c) | Trans1/cis1 | 34.7 | 44.1 | 33.8 | 14.7 | 4.2 |
| | Trans2/cis1 | 26.6 | 24.9 | 15.9 | 6.3 | 1.7 |

***Table S3.*** *The calculated ON/OFF ratio between trans and cis1 forms for the different scenarios.*